\documentclass[a4paper, 10pt]{article}
\usepackage{amsmath,amssymb,bm}
\usepackage[dvipdfm]{graphicx}
\title{Instabilities in associative memory model with synaptic depression and switching phenomena among attractors}
\author{Yosuke Otsubo$^{1\,\dagger}$,Kenji Nagata$^{1,2\,\ddagger}$, Masafumi Oizumi$^{1,2\,\dagger\dagger}$, Masato Okada$^{1,3\,\dagger\dagger\dagger}$
}
\begin{document}
\maketitle
\section*{Abstract}
We investigated how the stability of macroscopic states in the associative memory model is affected by synaptic depression. 
To this model, we applied the dynamical mean-field theory, which has recently 
been developed in stochastic neural network models with synaptic depression. 
By introducing a sublattice method, we derived macroscopic equations for firing state variables and depression variables. 
By using the macroscopic equations, we obtained the phase diagram when the strength of synaptic depression 
and the correlation level among stored patterns were changed.
We found that there is an unstable region in which both the memory state 
and mixed state cannot be stable and that various switching phenomena can occur in this region.
\footnote{
$^1$ {\it Graduate School of Frontier Sciences, The University of Tokyo, Kashiwa, Chiba 277-8561}\\
$^2$ {\it Research Fellow of the Japan Society for the Promotion of Science}\\
$^3$ {\it Brain Science Institute, RIKEN, Wako, Saitama, 351-0198}\\
$\dagger$ E-mail address: otsubo@mns.k.u-tokyo.ac.jp\\
$\ddagger$ E-mail address: nagata@mns.k.u-tokyo.ac.jp\\
$\dagger\dagger$ E-mail address: oizumi@mns.k.u-tokyo.ac.jp\\
$\dagger\dagger\dagger$E-mail address: okada@k.u-tokyo.ac.jp\\
}
\clearpage
\section{Introduction}
An associative memory model is one of typical neural network models 
that has discretely distributed fixed-point attractors as stored patterns.\cite{Nakano,Anderson,Kohonen, Hopfield, Mimura} \\
For this model, it is known that the macroscopic state of the network usually remains in an attractor. 
On the other hand, synaptic plasticity can destabilize the network and induce a periodic or 
aperiodic itinerancy of the macroscopic state. 
\cite{Pantic, Marro_1, Torres_2, Melamed}
Such switching phenomena, known as the dynamics among quasi-attractors, \cite{Sompolinsky,Tsuda, Amit, Fukai} have 
rich implications for the field of neurodynamics, e.g., 
communication from external world, memory search or symbol-emergence.\\
The switching phenomena can be observed by a model with synaptic depression, \cite{Pantic, Torres_2}
which is a physiological phenomenon in which high-frequency presynaptic inputs induce a decrease in synaptic weights.
\cite{Thomson,Abbott,Tsodyks}\\
In this study, we considered the associative memory model with synaptic depression. 
We considered correlated memory patterns as well as uncorrelated patterns as a general case. 
For this, we needed to treat not only the memory state but also the mixed state, which are both attractors.
The mixed state is the mixing of arbitrary memory patterns generated by correlation learning and is not simply a side effect that is
unnecessary for information processing.\cite{Amari,Toya} 
\\ 
Recently, Igarashi \textit{et al.} have proposed a dynamical mean-field theory of models with stochastic neurons\cite{Igarashi_0}. 
We applied this theory to the associative memory model 
and introduced the notion of a sublattice. 
By using the dynamical mean-field theory with sublattice method, 
we investigated how the strength of synaptic depression and the correlation level 
among stored patterns affect the stability of the memory and mixed states.
We found an unstable region in which both the memory and mixed states cannot be stable. 
In this region, there are various switching phenomena among the attractors.\\
The rest of this paper is organized as follows. 
Section 2 describes the model used in this paper. 
In \S 3, we discuss the theoretical method for the model in order to 
derive the macroscopic equations by the mean-field approach and the sublattice method, 
and we also introduce a 
stability analysis of the steady state. 
Section 4 presents results for how synaptic depression influences the stability and dynamics of 
the macroscopic  state according to the correlation level among stored patterns. 
In \S 5, we summarize the results presented in this paper.

\section{Model}
\label{sec:cls}
\begin{figure}
\begin{center}
\includegraphics[width=3.0in]{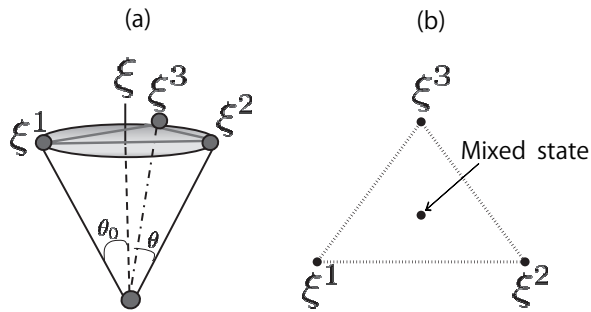}
\caption{(a)Schematic illustration of correlated memory patterns for $p=3$.
The direction cosines are $\cos\theta_0=b$ and $\cos\theta=b^2$ for eq. (\ref{eq:overlap_pattern}). 
(b) Relations among the memory patterns and the mixed state.}
\label{fig:result-1}
\end{center}
\end{figure}

We discuss an attractor network model with $N$ fully connected binary neurons.
If the $i$-th neuron fires at time $t$, its state is $s_i(t)=1$: otherwise, $s_i(t)=0$. 
Then, the state of the network is characterized by 
$\textrm{\boldmath $s$}(t)=(s_1(t), \cdots ,s_N(t))$.
The synaptic weight $J_{ij}(t)$ from presynaptic neuron $j$ to postsynaptic neuron $i$ at time $t$ 
changes dynamically owing to synaptic depression. 
We use synaptic weight $J_{ij}(t)$ incorporating the synaptic depression by a fixed synaptic weight $J_{ij}$
multiplied by a dynamic amplitude factor $x_j(t)$:
\begin{eqnarray}
\label{eq:interaction_1}
\lefteqn{}J_{ij}(t)&=&J_{ij}x_j(t),\\
x_j(t+1)&=&  x_j(t)+\frac{1-x_j(t)}{\tau}-Ux_j(t)s_j(t),
\label{eq:depression}
\end{eqnarray}
where $x_j(t)$ is determined by a phenomenological model of synapses,\cite{Abbott,Tsodyks,Pantic} 
and takes $0<x_j(t)\leq 1$, where $x_j(t)=1$ 
correspond to the case without synaptic depression.
The depression variable from the presyanptic neuron, $x_j(t+1)$, decreases by a certain fraction $Ux_j(t)$ after each spike is emitted, $s_j(t)=1$, and recovers with time constant $\tau$. \\
The system is simultaneously updated, i.e, the synchronous rule, and each neuron obeys 
probabilistic dynamics:
\begin{eqnarray}
\label{eq:glauber_dym}
\lefteqn{}&& \mathrm{Prob}[s_i(t+1)=1]=g_{\beta}\left( h_i(t)\right), \\
&& g_{\beta}(h)= \frac{1}{2}\left(1+\tanh \beta h \right),\nonumber
\end{eqnarray}
where $\beta=1/T$ represents the inverse temperature. 
The function $h_i (t)$ is the internal potential of the $i$-th neuron at time $t$, 
which is defined using the synaptic weight and the $i$-th neuron's state $s_i(t)$ as
\begin{equation}
h_i(t)=\sum_{j \neq i}J_{ij}(2s_j(t)x_j(t)-1).
\label{eq:h_original}
\end{equation}
In the case without synaptic depression, the system in eq. (\ref{eq:h_original}) 
returns to the well known Ising spin system with $\textrm{\boldmath $\sigma$}(t)=(\sigma_1(t),\cdots,\sigma_N(t))$, 
\begin{equation}
h_i(t)=\sum_{j\neq i}J_{ij}\sigma_j(t)\:\:(\sigma_j = \pm 1).
\label{eq:potential}
\end{equation}
We consider the associative memory model with correlated memory patterns. 
We introduce a parent pattern and $p$ child patterns. 
\begin{eqnarray}
\lefteqn{}\textrm{\boldmath $\xi$}=(\xi_1,\cdots,\xi_N),\:\:
\textrm{\boldmath $\xi$}^{\mu}=(\xi^{\mu}_1,\cdots,\xi^{\mu}_N),\:\:\:\mu=1,\cdots,p. 
\end{eqnarray}
These are random variables drawn from the following probability distributions:
\begin{equation}
\label{eq:xi_prob}
\mathrm{Prob}[\xi_i=\pm 1]=\frac{1}{2},\; \mathrm{Prob}[\xi_i^{\mu}= \pm 1]=\frac{1\pm b\xi_i}{2},
\end{equation}
where correlation coefficient $b$ takes $0\leq b\leq 1$ and 
represents the correlation level between stored patterns.
For $b=0$, child patterns are mutually orthogonal for $N\rightarrow \infty$; 
for $b=1$, they are the same as the parent pattern.
In this study, we treated child patterns as memory patterns, i.e., stored patterns, 
so at the thermodynamic limit of $N\rightarrow \infty$, 
the direction cosine between memory patterns can be described as
\begin{eqnarray}
\frac{1}{N}\sum_{i=1}^{N}\xi_i^{\mu}\xi_i^{\mu'}&=&\delta_{\mu\mu'}+b^2(1-\delta_{\mu\mu'}),
\label{eq:overlap_pattern}
\end{eqnarray}
where $\delta_{\mu\nu}$ is Kronecker's delta defined as
\begin{eqnarray}
\delta_{\mu\nu}=
\left\{
\begin{array}{ll}
1  &\quad (\mu=\nu) \\
0  &\quad (\mu \neq \nu).
\end{array}
\right.
\end{eqnarray}
The distance relationships among three stored patterns are shown in Fig. 1(a). 
In this example, the mixed state is defined as follows
\begin{equation}
\mathrm{sgn}(\textrm{\boldmath $\xi$}^1+\textrm{\boldmath $\xi$}^2+\textrm{\boldmath $\xi$}^3),
\end{equation}
where the output function $\mathrm{sgn}(\cdot)$ is
\begin{eqnarray}
\mathrm{sgn}(u)=
\left\{
\begin{array}{ll}
1  &\quad (u\geq 0) \\
-1  &\quad (u<0).
\end{array}
\right.
\end{eqnarray}
Figure \ref{fig:result-1}(b) shows a schematic diagram containing the mixed state, 
which corresponds to Fig. 1(a) viewed from above. 
The center dot in the triangle represents the mixed state.\\
The fixed synaptic weight $J_{ij}$ is set, according to the Hebbian rule, to
\begin{equation}
J_{ij}=\frac{1}{N}\sum_{\mu=1}^{p}\xi_i^{\mu}\xi_j^{\mu}.
\label{eq:interaction}
\end{equation}
A self-connection $J_{ii}$ is assumed not to exist.

\section{Analysis}
Since the synaptic weight $J_{ij}(t)=J_{ij}x_j(t)$ is asymmetric due to the dynamic amplitude factor $x_j(t)$,
we cannot treat the present system described by eqs. (\ref{eq:interaction_1})-(\ref{eq:potential})
by the conventional equilibrium statistical mechanical approach. 
In this section, we analyze the associative memory 
model with correlated memory patterns by a dynamical mean-field theory for finite temperature\cite{Igarashi_0}. 
Then, we 
derive the macroscopic steady-state equation 
by introducing the sublattice method.
\subsection{Mean-field analysis}
First, we consider the thermal average of 
the $i$-th neuron at time $t$, $\langle s_i(t) \rangle=\langle g_{\beta}(h_i(t)) \rangle$. 
Then, we get the following equations from eqs. (\ref{eq:glauber_dym}) and (\ref{eq:h_original}) 
by using the mean-field approximation,
\begin{equation}
\langle s_i(t) \rangle =g_{\beta}( \langle h_i(t) \rangle ),\: 
\langle h_i(t) \rangle =\sum_{j \neq i}J_{ij}(2 \langle s_j(t)x_j(t) \rangle -1), 
\label{eq:s_mean_angle}
\end{equation}
where $\langle \cdot \rangle$ denotes the thermal average with respect to eq. (\ref{eq:glauber_dym}).
Similarly, we take the thermal average of eq. (\ref{eq:depression}) for the dynamic amplitude factor,
\begin{equation}
\langle x_i(t+1)  \rangle=  \langle x_i(t) \rangle +\frac{1-\langle x_i(t) \rangle }{\tau}-U \langle s_i(t)x_i(t) \rangle.
\label{eq:x_mean_angle}
\end{equation}
If the number of memory patterns $p$ is on the order of $1$ with respect to the number of neurons $N$, 
the fixed synaptic weight $J_{ij}$ described by eq. (\ref{eq:interaction}) is on the order of $1/N$. 
In this case, the equal time correlation between $s_i(t)$ and $x_i(t)$ is on the order $1/N$. 
At the thermodynamic limit $N\rightarrow \infty$, we can consider $s_i(t)$ and $x_i(t)$ as being independent. 
The thermal average of the product of $s_i(t)$ and $x_i(t)$ can therefore be decoupled as\cite{Igarashi}
\begin{equation}
\langle s_i(t)x_i(t) \rangle=  \langle s_i(t) \rangle \langle x_i(t) \rangle.
\label{eq:mean_sep}
\end{equation}
By using this, we can rewrite 
eqs. (\ref{eq:s_mean_angle}) and (\ref{eq:x_mean_angle}), respectively, as
\begin{eqnarray}
\label{eq:s_mean_dym}
m_i(t+1)& =& g_{\beta}\left(\sum_{j\neq i }J_{ij}(2m_j(t)X_j(t)-1) \right),\\
\lefteqn{}X_i(t+1) & =&X_i(t)+\frac{1-X_i(t)}{\tau}-Um_i(t)X_i(t),
\label{eq:x_mean_dym}
\end{eqnarray}
where $m_i(t)\equiv\langle s_i(t) \rangle$ and $X_i(t) \equiv \langle x_i(t) \rangle $. 
Considering the steady states 
for the thermal average, $m_i \equiv m_i(\infty)$ and $X_i \equiv X_i(\infty)$, respectively, 
we obtain the following microscopic equation from eqs. (\ref{eq:s_mean_dym}) and (\ref{eq:x_mean_dym}),
\begin{equation}
m_i=g_{\beta}\left(\sum_{j \neq i}J_{ij} \left( \frac{2m_j}{1+\gamma m_j}-1 \right) \right),\: X_i=\frac{1}{1+\gamma m_i}, 
\label{eq:m_steady}
\end{equation}
where $\gamma \equiv \tau U$ indicates the level of synaptic depression in the steady state. 
For the above equations, the thermal average of the firing rate and the dynamic amplitude factor are 
determined by one depressing parameter $\gamma$ in the steady state.
Note that in previous studies, 
the correlation between the dynamic amplitude factor and the neuron's state was taken to be 
\begin{equation}
\langle s_ix_i \rangle =\left\langle \frac{s_i}{1+\gamma s_i} \right\rangle=\frac{m_i} {1+\gamma},
\end{equation}
since $s_i$ takes a binary value, $s_i=\left\{0,1 \right\}$, 
at a low temperature $T\sim0\: (1\ll\beta<\infty)$.\cite{Matsumoto} \cite{Mejias}
By contrast, we extend the correlation to the case of finite temperature in the form 
\begin{equation}
\langle s_ix_i \rangle =m_iX_i =\frac{m_i}{1+\gamma m_i}.
\label{eq:mx}
\end{equation}

\subsection{Macroscopic steady-state equation}
In this section, we obtain the macroscopic steady-state equations 
from the microscopic ones described by eq. (\ref{eq:m_steady}) in terms of 
the sublattice notion. 
This method is essential to describe the macroscopic state of the network \cite{Torres}.
By substituting the Hebbian rule of eq. (\ref{eq:interaction}) into eq. (\ref{eq:m_steady}), 
we derive the following microscopic equation.
\begin{equation}
 m_i=g_{\beta}\left( \frac{1}{N}\sum_{\mu=1}^{p}\xi_i^{\mu}\sum_{j \neq i}\xi_{j}^{\mu}\left(\frac{2m_j}{1+\gamma m_j}-1 \right) \right).
\end{equation}
To obtain the macroscopic steady-state equation, we define $p$ dimensional memory patterns 
$\textrm{\boldmath $\xi$}_i=(\xi_i^1,\cdots,\xi_i^p)^{T}\in \left\{-1,1 \right\}^p$, 
where the superscript $^T$ stands for transposition.
On the basis of these vectors, a set of neurons $\left\{1,\cdots,N \right\}$ is divided into $2^p$ groups as
\begin{equation}
\mathcal{I}_{\textrm{\boldmath $\eta$}}
=\left\{ i|\textrm{\boldmath $\xi$}_i=\textrm{\boldmath $\eta$} \right\},  \quad  
\left\{1,\cdots,N \right\}=\bigcup _{\textrm{\boldmath $\eta$}}\mathcal{I}_{\textrm{\boldmath $\eta$}},
\end{equation}
where $\mathcal{I}_{\textrm{\boldmath $\eta$}}$ is called a sublattice and 
$\textrm{\boldmath $\eta$} =(\eta^1,{\eta}^2,...,{\eta}^p)^{T}\in \left\{-1,1 \right\}^p $.\cite{Coolen}
For example, in the case of three memory patterns divided into eight sublattices, we can introduce the following combination.
\begin{equation}
\textrm{\boldmath $\eta$}
=
\left(
\begin{array}{c}
1 \\
1\\
1
\end{array}
\right),
\left(
\begin{array}{c}
1 \\
1\\
-1
\end{array}
\right),
\left(
\begin{array}{c}
1 \\
-1\\
1
\end{array}
\right),
\left(
\begin{array}{c}
1 \\
-1\\
-1
\end{array}
\right),
\left(
\begin{array}{c}
-1 \\
1\\
1
\end{array}
\right),
\left(
\begin{array}{c}
-1 \\
1\\
-1
\end{array}
\right),
\left(
\begin{array}{c}
-1 \\
-1\\
1
\end{array}
\right),
\left(
\begin{array}{c}
-1 \\
-1\\
-1
\end{array}
\right).
\label{eq:memory_mat}
\end{equation}
Because the memory patterns are 
produced by eq.(\ref{eq:xi_prob}),
the number of neurons $|\mathcal{I}_{\textrm{\boldmath $\eta$}}|$ in the sublattice $\mathcal{I}_{\textrm{\boldmath $\eta$}}$ is 
\begin{eqnarray}
|\mathcal{I}_{\textrm{\boldmath $\eta$}}|=
\left\{
\begin{array}{ll}
\frac{b_{+}^3+b_{-}^3}{2}N,  &\quad  \mathrm{if}\:\textrm{\boldmath $\eta$}=(1,1,1)^{T},(-1,-1,-1)^{T}\\
\frac{b_{+}b_{-}}{2}N,  &\quad \mathrm{otherwise},
\end{array}
\right.
\end{eqnarray}
where $b_{\pm}\equiv \frac{1\pm b}{2}$ 
and $|\mathcal{I}_{\textrm{\boldmath $\eta$}}|$ is $O(N)$ since $2^p$ is $O(1)$.\\
Following the expression of the sublattice, we can rewrite the fixed synaptic weight (\ref{eq:interaction}) as 
\begin{eqnarray}
\lefteqn{}J_{ij}&=&\frac{1}{N}\sum_{\mu=1}^{p}\eta^{\mu}\eta'^{\mu}\:\:
(i\in \mathcal{I}_{\textrm{\boldmath $\eta$}}, j\in \mathcal{I}_{\textrm{\boldmath $\eta$}'})\\
 &=& \frac{1}{N}\textrm{\boldmath $\eta$}\cdot\textrm{\boldmath $\eta$}'.
\label{eq:sub_interaction}
\end{eqnarray}
We can regard the model in this study as an extended Hushimi-Temperly (HT) model 
because the synaptic weight between neurons within the same sublattice is constant, i.e., $J_{ij}=\frac{p}{N}$. 
Since the synaptic weight between neurons is determined by which sublattice they belong to 
(see eq. (\ref{eq:sub_interaction})),
the firing rate, the internal potential, and dynamic amplitude factor of neurons within the same sublattice are the same.
Therefore, we can introduce the sublattice firing rate $m_{\textrm{\boldmath $\eta$}}$, 
internal potential $h_{\textrm{\boldmath $\eta$}}$, and dynamic amplitude factor $X_{\textrm{\boldmath $\eta$}}$ as 
$m_{\textrm{\boldmath $\eta$}}\equiv m_i=m_j$, $h_{\textrm{\boldmath $\eta$}}\equiv h_i=h_j$, 
and $X_{\textrm{\boldmath $\eta$}}\equiv X_i=X_j$ 
on the condition that $i\in \mathcal{I}_{\textrm{\boldmath $\eta$}}$ and $j\in \mathcal{I}_{\textrm{\boldmath $\eta$}}$.\\
By eqs. (\ref{eq:m_steady}) and (\ref{eq:sub_interaction}), 
the macroscopic steady-state equations are given by: 
\begin{eqnarray}
\label{eq:sub_steady}
m_{\textrm{\boldmath $\eta$}} &=&g_{\beta}( \langle h_{\textrm{\boldmath $\eta$}} \rangle ),\:X_{\textrm{\boldmath $\eta$}}
=\frac{1}{1+\gamma m_{\textrm{\boldmath $\eta$}}},\\
\langle h_{\textrm{\boldmath $\eta$}}\rangle&=&\sum_{\textrm{\boldmath $\eta$}'}p_{\textrm{\boldmath $\eta$}'}
\textrm{\boldmath $\eta$} 
\cdot\textrm{\boldmath $\eta$}'
 \left( \frac{2m_{\textrm{\boldmath $\eta$}'}}{1+\gamma m_{\textrm{\boldmath $\eta$}'}}-1 \right),
\end{eqnarray}
where $ p_{\textrm{\boldmath $\eta$}}\equiv \frac{|\mathcal{I}_{\textrm{\boldmath $\eta$}}|}{N} $ denotes the relative sublattice size.\\
Next, we introduce the closeness between the state of system $\textrm{\boldmath $s$}(t)$ at time $t$ and 
the $\mu$-th memory pattern $\textrm{\boldmath $\xi$}^{\mu}$ characterized by an overlap
\begin{equation}
M^{\mu}(t)=\frac{1}{N}\sum_{i=1}^N\xi^{\mu}_i(2s_i(t)-1).
\label{eq:overlap}
\end{equation}
By following sublattice method, we can also describe the above equation in the steady state as
\begin{equation}
M^{\mu}=\sum_{\textrm{\boldmath $\eta$}'}p_{\textrm{\boldmath $\eta$}'}{\eta'}^{\mu}(2m_{\textrm{\boldmath $\eta$}'}-1).
\label{eq:sub_overlap}
\end{equation}
If state $\textrm{\boldmath $s$}(t)$ corresponds to memory pattern $\textrm{\boldmath $\xi$}^{\mu}$, 
then $M^{\mu}(t)$ 
is exactly $1$ at $N\rightarrow \infty$. 
The purpose of the sublattice method is to treat the macroscopic variables, 
$2^p$ sublattices, 
instead of microscopic variables, $N$ neurons, by 
grouping homogeneous neurons with respect to memory patterns $\textrm{\boldmath $\xi$}_i=(\xi_i^1,...,\xi_i^p)$.

\subsection{Stability analysis of macroscopic steady state}
In this section, we discuss the stability of eqs. (\ref{eq:s_mean_dym}) and (\ref{eq:x_mean_dym}). 
If the neurons belong to the same sublattice, we consider the steady state of the neurons to be the same state.
First, in order to make a correspondence between the neuron index and sublattice index, 
we relabel the $i$-th neuron in the sublattice $\mathcal{I}_{\textrm{\boldmath $\eta$}}$ 
using index $l$,
\begin{eqnarray}
i\: \rightarrow \: (\textrm{\boldmath $\eta$},l),\:\:l=1,\cdots ,|\mathcal{I}_{\textrm{\boldmath $\eta$}}|.
\end{eqnarray}
Under this mapping, the time-dependent firing rate and dynamic amplitude factor can be expressed by 
\begin{eqnarray}
m_i (t)\: \rightarrow m_l^{\textrm{\boldmath $\eta$}}(t),\\
X_i (t)\: \rightarrow X_l^{\textrm{\boldmath $\eta$}}(t).
\end{eqnarray} 
Next, we rewrite the above functions as 
\begin{eqnarray}
m_l^{\textrm{\boldmath $\eta$}}(t)&=&m_{\textrm{\boldmath $\eta$}}+\delta m_l^{\textrm{\boldmath $\eta$}}(t),\\
X_l^{\textrm{\boldmath $\eta$}}(t)&=&X_{\textrm{\boldmath $\eta$}}+\delta X_l^{\textrm{\boldmath $\eta$}}(t),
\end{eqnarray}
where $\:\delta m_l^{\textrm{\boldmath $\eta$}}(t)$ and $\delta X_l^{\textrm{\boldmath $\eta$}}(t)$ 
denote the small deviations around steady point $m_{\textrm{\boldmath $\eta$}}$ and $X_{\textrm{\boldmath $\eta$}}$ respectively.
Here, if the neurons belong to the same sublattice, we consider their steady states to be the same. 
Form eqs. (\ref{eq:s_mean_dym}), (\ref{eq:x_mean_dym}), and (\ref{eq:sub_interaction}), these fluctuations are 
\begin{eqnarray}
\label{eq:m_dym_am}
\delta m_l^{\textrm{\boldmath $\eta$}}(t+1)=4 \beta m_{\textrm{\boldmath $\eta$}}
\left( 1-m_{\textrm{\boldmath $\eta$}} \right)\frac{1}{N}
\sum_{\textrm{\boldmath $\eta$}}\sum_{l'=1}^{|\mathcal{I}_{\textrm{\boldmath $\eta$}'}|}
\sum_{\mu=1}^{p}
\eta^{\mu}\eta'^{\mu}
(X_{\textrm{\boldmath $\eta$}'}\delta m_{l'}^{\textrm{\boldmath $\eta$}'}(t)
+m_{\textrm{\boldmath $\eta$}'}\delta X_{l'}^{\textrm{\boldmath $\eta$}'}(t)),\: (l' \in \mathcal{I}_{\textrm{\boldmath $\eta$}'}),\\
\delta X_l^{\textrm{\boldmath $\eta$}}(t+1)
=
-UX_{\textrm{\boldmath $\eta$}}\delta m_l^{\textrm{\boldmath $\eta$}}(t)
+\left(1-\frac{1}{\tau}-Um_{\textrm{\boldmath $\eta$}}\right)\delta X_l^{\textrm{\boldmath $\eta$}}(t),
\label{eq:x_dym_am}
\end{eqnarray}
where we use the relation $\sum_{j\neq i}h_j(t)\sim  \sum_{j=1}^{N}h_{j}(t)$ for $J_{ij}\sim O(1/N)$. 
The sequence of small deviations, $\left\{\delta m_l^{\textrm{\boldmath $\eta$}} \right\}$ and $\left\{\delta X_l^{\textrm{\boldmath $\eta$}} \right\}$, 
can be rewritten by Fourier transformation as follows: 
\begin{eqnarray}
\label{eq:fou2}
\delta \hat{m}_k^{\textrm{\boldmath $\eta$}}(t)&=&\frac{1}{|\mathcal{I}_{\textrm{\boldmath $\eta$}}|}\sum_{l} 
\delta  {m}_{l}^{\textrm{\boldmath $\eta$}}(t) 
e^{-2\pi \mathrm{i}k l/|\mathcal{I}_{\textrm{\boldmath $\eta$}}|},\\
\delta \hat{X}_k^{\textrm{\boldmath $\eta$}}(t)&=&\frac{1}{|\mathcal{I}_{\textrm{\boldmath $\eta$}}|}\sum_{l} 
\delta  {X}_{l}^{\textrm{\boldmath $\eta$}}(t) 
e^{-2\pi \mathrm{i}k l/|\mathcal{I}_{\textrm{\boldmath $\eta$}}|}.
\end{eqnarray}
And the inverse Fourier transformations are given by 
\begin{eqnarray}
\label{eq:fou}
\delta m_l^{\textrm{\boldmath $\eta$}}(t)&=&\sum_{k} 
\delta \hat {m}_{k}^{\textrm{\boldmath $\eta$}}(t) 
e^{2\pi \mathrm{i}k l/|\mathcal{I}_{\textrm{\boldmath $\eta$}}|},\\
\delta X_l^{\textrm{\boldmath $\eta$}}(t)&=&\sum_{k} 
\delta \hat {X}_{k}^{\textrm{\boldmath $\eta$}}(t) 
e^{2\pi \mathrm{i}k l/|\mathcal{I}_{\textrm{\boldmath $\eta$}}|}.
\end{eqnarray}
By using these equations, we can rewrite the dynamics of these functions described by eqs. (\ref{eq:m_dym_am}) and (\ref{eq:x_dym_am}) 
in the following form.
\begin{eqnarray}
\label{eq:am_deltam_f}
\sum_{k} \delta \hat {m}_{k}^{\textrm{\boldmath $\eta$}}(t+1) e^{2\pi \mathrm{i}k l/|\mathcal{I}_{\textrm{\boldmath $\eta$}}|}&=&
4 \beta m_{\textrm{\boldmath $\eta$}}(1-m_{\textrm{\boldmath $\eta$}})\frac{1}{N}
\sum_{\textrm{\boldmath $\eta$}'}\sum_{l'=1}^{|\mathcal{I}_{\textrm{\boldmath $\eta$}'}|}
\sum_{\mu=1}^{p}\sum_{k'}
\eta^{\mu}\eta'^{\mu}\nonumber\\
&& \times(X_{\textrm{\boldmath $\eta$}'}
\delta \hat{m}^{\textrm{\boldmath $\eta$}'}_{k'}(t)+
m_{\textrm{\boldmath $\eta$}'}\delta \hat{X}^{\textrm{\boldmath $\eta$}'}_{k'}(t))e^{2\pi \mathrm{i}k' l'/|\mathcal{I}_{\textrm{\boldmath $\eta$}'}|} ,
\end{eqnarray}
\begin{eqnarray}
\sum_{k} \delta \hat {X}_{k}^{\textrm{\boldmath $\eta$}}(t+1) e^{2\pi \mathrm{i}k l/|\mathcal{I}_{\textrm{\boldmath $\eta$}}|}=
\sum_{k'} 
\left\{ -UX_{\textrm{\boldmath $\eta$}} \delta \hat{m}_{k'}^{\textrm{\boldmath $\eta$}}(t)+
\left(1-\frac{1}{\tau}-U{m}_{\textrm{\boldmath $\eta$}}\right) \delta \hat{X}_{k'}^{\textrm{\boldmath $\eta$}}(t) \right\}
e^{2\pi \mathrm{i}k' l/|\mathcal{I}_{\textrm{\boldmath $\eta$}}|} ,&&
\label{eq:am_deltax_f}
\end{eqnarray}
where $\mathrm{i}$ is the square root of $-1$.\\
Since Fourier components are orthonormal, we obtain the following equations by comparing the 0-th coefficients 
in the above equations: 
\begin{eqnarray}
\delta \hat {m}_{0}^{\textrm{\boldmath $\eta$}}(t+1)&=&
4 \beta m_{\textrm{\boldmath $\eta$}}(1-m_{\textrm{\boldmath $\eta$}})\frac{1}{N}
\sum_{\textrm{\boldmath $\eta$}}\sum_{l'=1}^{|\mathcal{I}_{\textrm{\boldmath $\eta$}'}|}
\sum_{\mu=1}^{p}
\eta^{\mu}\eta'^{\mu}
(X_{\textrm{\boldmath $\eta$}'}\delta \hat{m}^{\textrm{\boldmath $\eta$}'}_{0}(t)
+m_{\textrm{\boldmath $\eta$}'}\delta \hat{X}^{\textrm{\boldmath $\eta$}'}_{0}(t))\\
&=&
\label{eq:m_0_dym}
4 \beta m_{\textrm{\boldmath $\eta$}}(1-m_{\textrm{\boldmath $\eta$}})\sum_{\textrm{\boldmath $\eta$}'}p_{\textrm{\boldmath $\eta$}'}
\textrm{\textrm{\boldmath $\eta$}}\cdot\textrm{\textrm{\boldmath $\eta$}}'
(X_{\textrm{\boldmath $\eta$}'}\delta \hat{m}^{\textrm{\boldmath $\eta$}'}_{0}(t)
+m_{\textrm{\boldmath $\eta$}'}\delta \hat{X}^{\textrm{\boldmath $\eta$}'}_{0}(t)),\\
\delta \hat {X}_{0}^{\textrm{\boldmath $\eta$}}(t+1)
&=&
-UX_{\textrm{\boldmath $\eta$}} \delta \hat {m}_{0}^{\textrm{\boldmath $\eta$}}(t)+
\left(1-\frac{1}{\tau}-U{m}_{\textrm{\boldmath $\eta$}}\right) \delta \hat {X}_{0}^{\textrm{\boldmath $\eta$}}(t).
\label{eq:x_0_dym}
\end{eqnarray}
Therefore, we get the following equation written in matrix form.
\begin{equation}
\left(
\begin{array}{c}
 \delta \hat{m}_0^{^{(+1,\cdots,+1)^T}} (t+1) \\
  \vdots\\
 \delta \hat{m}_0^{^{( -1,\cdots,-1)^T}}(t+1) \\
 \delta \hat{X}_0^{^{(+1,\cdots,+1)^T}} (t+1)\\
  \vdots\\
 \delta \hat{X}_0^{^{(-1,\cdots,-1)^T}}(t+1)\\
\end{array}
\right)
=\tilde{H}
\left(
\begin{array}{c}
\delta \hat{m}_0^{^{(+1,\cdots,+1)^T}} (t) \\
  \vdots \\
 \delta \hat{m}_0^{^{(-1,\cdots,-1)^T}}(t) \\
 \delta \hat{X}_0^{^{(+1,\cdots,+1)^T}} (t)\\
  \vdots \\
 \delta \hat{X}_0^{^{(-1,\cdots,-1)^T}}(t)\\
\end{array}
\right),
\end{equation}
where $\tilde{H}$ is a $2^{p+1} \times 2^{p+1}$ matrix consisting of four $2^p \times 2^p$ block matrices 
A, B, C, and D.
\begin{eqnarray}
\label{eq:hessian}
\tilde{H}&=&\left(
\begin{array}{cc}
A &B  \\
 C&D 
\end{array}
\right).
\end{eqnarray}
From eqs. (\ref{eq:m_0_dym}) and (\ref{eq:x_0_dym}), the elements of the $2^p \times 2^p$ block matrices are given by
\begin{eqnarray}
A_{{\textrm{\boldmath $\eta$}}{\textrm{\boldmath $\eta$}'}}&=&
4\beta {m}_{\textrm{\boldmath $\eta$}} (1-{m}_{\textrm{\boldmath $\eta$}})p_{\textrm{\boldmath $\eta$}'}\textrm{\boldmath $\eta$}\cdot\textrm{\boldmath $\eta$}'
X_{\textrm{\boldmath $\eta$}'},\\
B_{{\textrm{\boldmath $\eta$}}{\textrm{\boldmath $\eta$}'}}&= &
4\beta {m}_{\textrm{\boldmath $\eta$}} (1-{m}_{\textrm{\boldmath $\eta$}})p_{\textrm{\boldmath $\eta$}'}\textrm{\boldmath $\eta$}\cdot\textrm{\boldmath $\eta$}'
m_{\textrm{\boldmath $\eta$}'},\\ 
\label{eq:hessian_xm}
C_{{\textrm{\boldmath $\eta$}}{\textrm{\boldmath $\eta$}'}}& =&-U{X}_{\textrm{\boldmath $\eta$}}\delta_{{\textrm{\boldmath $\eta$}}{\textrm{\boldmath $\eta$}'}}, \\
D_{{\textrm{\boldmath $\eta$}}{\textrm{\boldmath $\eta$}'}}& =&\left(1-\frac{1}{\tau}-U{m}_{\textrm{\boldmath $\eta$}}\right)\delta_{{\textrm{\boldmath $\eta$}}{\textrm{\boldmath $\eta$}'}}. 
\label{eq:hessian_xx}
\end{eqnarray}
Next, by considering the $k$-th coefficient in eq. (\ref{eq:am_deltam_f}), we obtain:
\begin{eqnarray}
\nonumber
\delta \hat {m}_{k}^{\textrm{\boldmath $\eta$}}(t+1) e^{2\pi \mathrm{i}k l/|\mathcal{I}_{\textrm{\boldmath $\eta$}}|}&=&
4 \beta m_{\textrm{\boldmath $\eta$}}(1-m_{\textrm{\boldmath $\eta$}})\frac{1}{N}\sum_{\textrm{\boldmath $\eta$}'}
\sum_{l'=1}^{|\mathcal{I}_{\textrm{\boldmath $\eta$}'}|}\sum_{\mu=1}^{p}\eta^{\mu}\eta'^{\mu}\nonumber \\
&&\times (X_{\textrm{\boldmath $\eta$}'}\delta \hat{m}^{\textrm{\boldmath $\eta$}'}_{k}(t)
+m_{\textrm{\boldmath $\eta$}'}\delta \hat{X}^{\textrm{\boldmath $\eta$}'}_{k}(t))
e^{2\pi \mathrm{i}k l'/|\mathcal{I}_{\textrm{\boldmath $\eta$}'}|}\\ 
&=&
4 \beta m_{\textrm{\boldmath $\eta$}}(1-m_{\textrm{\boldmath $\eta$}})\frac{1}{N}\sum_{\textrm{\boldmath $\eta$}'}
\textrm{\textrm{\boldmath $\eta$}}\cdot\textrm{\textrm{\boldmath $\eta$}}' \nonumber \\
&&\times(X_{\textrm{\boldmath $\eta$}'}\delta \hat{m}^{\textrm{\boldmath $\eta$}'}_{k}(t)+
m_{\textrm{\boldmath $\eta$}'}\delta \hat{X}^{\textrm{\boldmath $\eta$}'}_{k}(t))
\sum_{l'=1}^{|\mathcal{I}_{\textrm{\boldmath $\eta$}'}|}e^{2\pi \mathrm{i}k l'/|\mathcal{I}_{\textrm{\boldmath $\eta$}'}|}.
\end{eqnarray}
Here, by eq. (\ref{eq:fou}), the following relation holds.
\begin{equation}
\sum_{l'=1}^{|\mathcal{I}_{\textrm{\boldmath $\eta$}'}|}e^{2\pi \mathrm{i}k l'/|\mathcal{I}_{\textrm{\boldmath $\eta$}'}|}=0,\:(k\neq 0).
\end{equation}
Consequently, the higher-order coefficients of eqs. (\ref{eq:am_deltam_f}) and (\ref{eq:am_deltax_f}) are 
\begin{eqnarray}
\label{eq:m_high}
\delta \hat {m}_{k}^{\textrm{\boldmath $\eta$}}(t+1)&=&0,\\
\delta \hat {X}_{k}^{\textrm{\boldmath $\eta$}}(t+1)
&=&
-UX_{\textrm{\boldmath $\eta$}} \delta \hat {m}_{k}^{\textrm{\boldmath $\eta$}}(t)
+\left(1-\frac{1}{\tau}-U{m}_{\textrm{\boldmath $\eta$}}\right) \delta \hat {X}_{k}^{\textrm{\boldmath $\eta$}}(t),
\label{eq:x_high}
\end{eqnarray}
and the above equations are represented in matrix form as 
\begin{equation}
\left(
\begin{array}{c}
 \delta \hat{m}_k^{^{(+1,\cdots,+1)^T}} (t+1) \\
  \vdots\\
 \delta \hat{m}_k^{^{( -1,\cdots,-1)^T}}(t+1) \\
 \delta \hat{X}_k^{^{(+1,\cdots,+1)^T}} (t+1)\\
  \vdots\\
 \delta \hat{X}_k^{^{(-1,\cdots,-1)^T}}(t+1)\\
\end{array}
\right)
=\tilde{H}'
\left(
\begin{array}{c}
\delta \hat{m}_k^{^{(+1,\cdots,+1)^T}} (t) \\
  \vdots \\
 \delta \hat{m}_k^{^{(-1,\cdots,-1)^T}}(t) \\
 \delta \hat{X}_k^{^{(+1,\cdots,+1)^T}} (t)\\
  \vdots \\
 \delta \hat{X}_k^{^{(-1,\cdots,-1)^T}}(t)\\
\end{array}
\right).
\end{equation}
From eqs. (\ref{eq:m_high}) and (\ref{eq:x_high}), 
the $2^{p+1}\times 2^{p+1}$ matrix $\tilde{H}'$ 
can be written as
\begin{eqnarray}
\tilde{H}'&=&\left(
\begin{array}{cc}
0 &0 \\
 C&D
\end{array}
\right),
\end{eqnarray}
where the elements of each block matrix, $C$ and $D$, are given by the same form 
of eqs. (\ref{eq:hessian_xm}) and (\ref{eq:hessian_xx}), respectively.
The eigenvalues of $\tilde{H}'$ are
\begin{equation}
\lambda_{\textrm{\boldmath $\eta$}}=0,\:1-\frac{1}{\tau}-Um_{\textrm{\boldmath $\eta$}}.
\end{equation}
Since the above eigenvalues do not exceed 1 for $\tau \geq 1$, $0<U\leq1$ and $0 \leq m_{\textrm{\boldmath $\eta$}} \leq 1$, 
the higher-order components do not influence the stability of the steady-state solution.
Therefore, we need to investigate only the eigenvalues of the Hessian matrix described by eq. (\ref{eq:hessian}) 
in order to obtain the stability of the steady states. 
In other words, if the maximum eigenvalue $|\lambda|_{max}$ of $\tilde{H}$ satisfies the condition 
$|\lambda|_{max}<1$, the macroscopic steady state described by eq. (\ref{eq:sub_steady}) is stable.

\section{Results}
In this section, we present the results of our investigation of the macroscopic behavior of 
the associative memory model with synaptic depression. 
We considered the associative memory model embedded with correlated memory patterns described by eq. (\ref{eq:memory_mat}).

\begin{figure}[t]
\begin{center}
\includegraphics[width=4.6in]{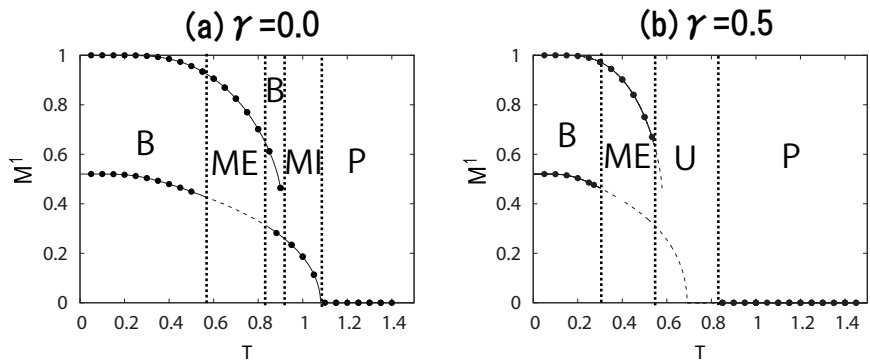}
\caption{Dependence of overlap $M^{1}$ on finite temperature $T$ at $\tau=100.0$ and $b=0.2$. 
(a) Without synaptic depression, i.e., $\gamma=0.0$. 
(b) With synaptic depression at $\gamma=0.5$. 
Solutions on solid lines are stable, while those on dashed lines are unstable. 
Black dots were obtained by computer simulation with $N=9.6 \times 10^4$.}
\label{fig:M-T}
\end{center}
\end{figure}

\subsection{Stability analysis for finite temperature}
\begin{figure}[t]
\begin{center}
\includegraphics[width=4.5in]{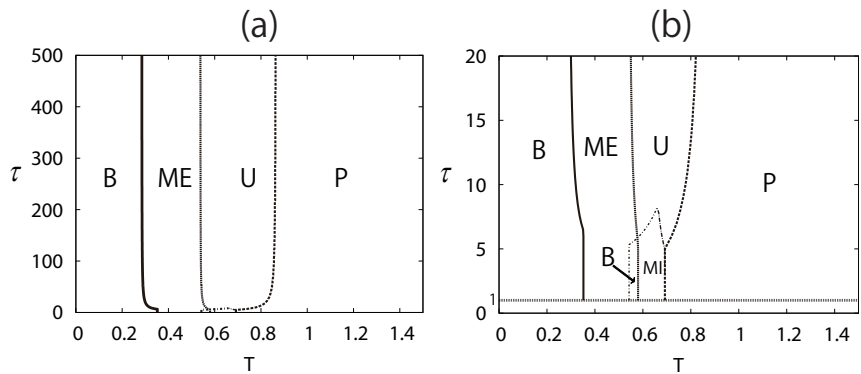}
\caption{Phase diagram for stability in parameter $(T,\tau)$ space. 
(a) For a broad array of time constant $\tau$ at $\gamma=0.5$ and $b=0.2$. 
(b) Magnified view lower $\tau$-region of (a) 
showing where the transition of the unstable phase begins. 
 Each line represents a phase boundary.}
\label{fig:tau_T}
\end{center}
\end{figure}

We examined the stability of the steady state for finite temperature 
by numerical analysis using eq. (\ref{eq:sub_steady}) under various conditions.\\
There are three kinds of solutions to eq. (\ref{eq:sub_steady}), 
which are called the ``memory state", ``mixed state", and ``paramagnetic state". 
The memory state corresponds to the state near a memory pattern.
In this state, for example, the overlap is represented by $(M^1,M^2,M^3)=(M,M',M')$ with $|M|>|M'|$.
The mixed state corresponds to the state that is in the center among memory patterns, i.e.,
$\mathrm{sgn}(\textrm{\boldmath $\xi$}^1+\textrm{\boldmath $\xi$}^2+\textrm{\boldmath $\xi$}^3)$.
In this state, the overlap of every memory pattern is the same, i.e., $(M^1,M^2,M^3)=(M,M,M)$. 
In the paramagnetic state, each neuron's state is random. 
Thus, the overlap of every memory pattern is 0, i.e., $(M^1,M^2,M^3)=(0,0,0)$. \\
Figures \ref{fig:M-T}(a) and \ref{fig:M-T}(b) show how the overlap $M^{1}$ defined by eq. (\ref{eq:overlap})
depended on temperature $T$ at $\tau=100.0$ and $b=0.2$.
We considered only the case of $M^1\geq 0$ in these figures, 
because the overlap was symmetric between positive and negative. 
The solutions on the solid lines are stable and those on the dashed lines are unstable. 
The black dots represent the numerical result obtained by computer simulation with $N=9.6 \times 10^4$.
There is good agreement between the stable solution obtained from eq. (27) (solid line)
and computer simulation of eq. (3) (black dots). 
This indicates that the framework of the sublattice method is appropriate for describing the macroscopic state. \\
The value $\gamma$ was fixed at $0.0$ in Fig. \ref{fig:M-T}(a) and at $0.5$ in Fig. \ref{fig:M-T}(b). 
The former represents the case without synaptic depression. 
We can divide the results in Fig. \ref{fig:M-T}(a) into five phases on the basis of the solutions of eq. (\ref{eq:sub_steady}). 
In the memory state phase, denoted by ``ME", only the solutions of the memory state were stable. 
In the mixed state phase, denoted by ``MI", only the solutions of the mixed state were stable. 
At a low temperature, there was bistable phase ``B", in which both the memory and mixed states could be stable. 
At a high temperature, the state went into paramagnetic phase ``P".\\
On the other hand, there is region ``U" in Fig. \ref{fig:M-T}(b). 
We call this region an unstable phase. 
In this phase, there was no steady state. 
By comparing Figs. \ref{fig:M-T}(a) and \ref{fig:M-T}(b), we see that the unstable phase
arose from the effect of synaptic depression at intermediate temperatures. 
As described in the next section, interesting behavior of the macroscopic states could be observed in the unstable phase. \\
In this section, we show phase diagrams when the depression time constant $\tau$ and memory pattern correlation coefficient $b$ 
were changed together with temperature $T$. 
In particular, we investigated in what parameter region unstable phase occurred. \\
First, we show the phase diagram of the macroscopic state with respect to $\tau$ at $\gamma=0.5$ and $b=0.2$ in
Figs. \ref{fig:tau_T}(a) and \ref{fig:tau_T}(b).\\ 
\begin{figure}[t]
\begin{center}
\includegraphics[width=3.0in]{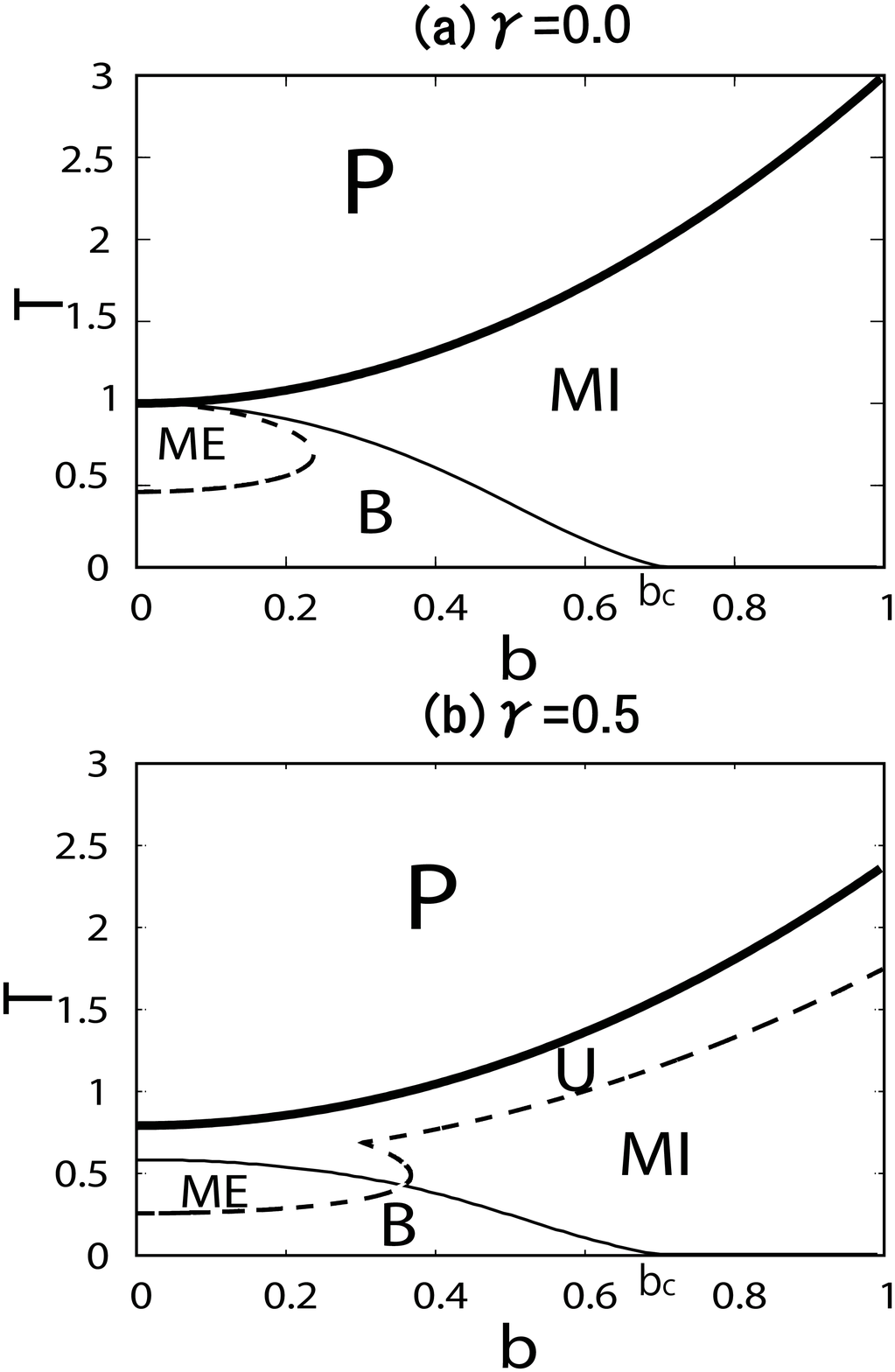}
\caption{Phase diagram for stability in parameter $(b,T)$ space. 
(a) Without synaptic depression, i.e., $\gamma=0.0$. 
(b) With synaptic depression at $\gamma=0.5$ and $\tau=100.0$. 
Thick solid lines represent phase boundaries of the paramagnetic phase. 
Thin and dashed lines represent 
the phase boundaries of the memory and 
mixed state phases, respectively.}
\label{fig:b-T}
\end{center}
\end{figure}
\begin{figure}[t]
\begin{center}
\includegraphics[width=4.0in]{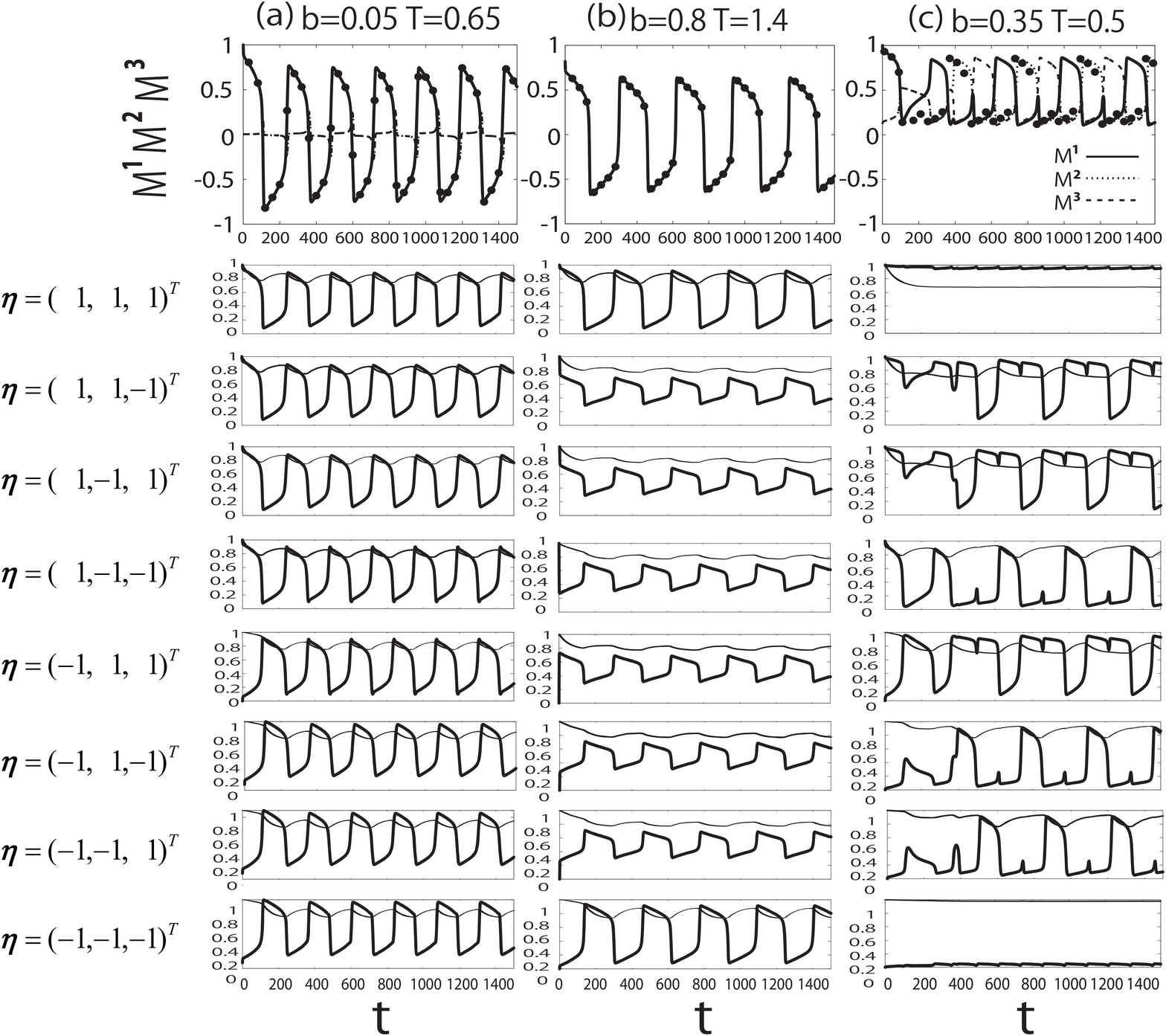}
\caption{
Periodic oscillatory behavior in the unstable phase at $\gamma=0.5$. 
The top figure shows the dependence of each overlap $M^1$, $M^2$ and $M^3$ on time $t$ 
and the bottom eight figures show
the dependences of sublattice firing rate $m_{\textrm{\boldmath $\eta$}}$ and 
dynamic amplitude factor $X_{\textrm{\boldmath $\eta$}}$ 
on time $t$ 
at (a) $b=0.05$, $T=0.65$, (b) $b=0.8$, $T=1.4$ and (c) $b=0.35$, $T=0.5$.
}
\label{fig:osci}
\end{center}
\end{figure}

\begin{figure}[t]
\begin{center}
\includegraphics[width=4.0in]{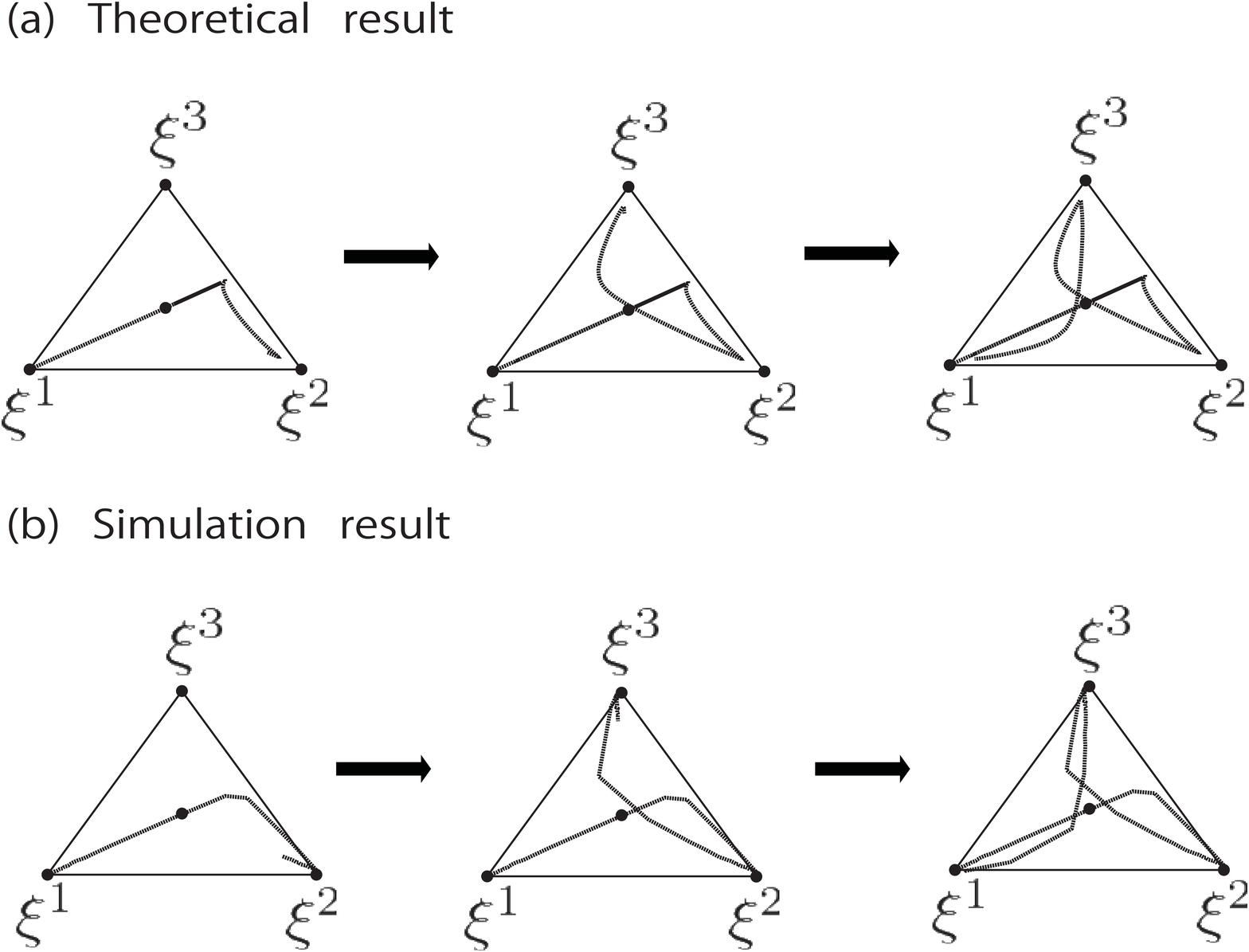}
\caption{Dynamical process of a macroscopic state 
corresponding to Fig. \ref{fig:osci}(c) 
in phase space composed of overlaps $(M^1,M^2,M^3)$. 
(a) Theoretical results. (b) Simulation results.
In this space, memory patterns, $\xi^1$, $\xi^2$, and $\xi^3$ correspond to $(1,b^2,b^2)$, $(b^2,1,b^2)$, 
and $(b^2,b^2,1)$, respectively. 
The center dots in the triangles represent the mixed state. The solid lines in the triangles 
indicate the trajectory of the network state.}
\label{fig:osci_2}
\end{center}
\end{figure}

Figure \ref{fig:tau_T}(a) shows the phase diagram at larger $\tau$, 
while Fig. \ref{fig:tau_T}(b) shows it at smaller $\tau$. 
In these results, we can see that each transition temperature was constant 
when $\tau$ was sufficiently large. 
Furthermore, there was no unstable phase at a small $\tau$. 
Here, note that our model is not valid for $\tau<1$. 
Even though parameter $\gamma$ was constant, the stability of the network could 
change depending on time constant $\tau$. \\
Next, we show the phase diagram for stability according to correlation coefficient $b$ in Figs. \ref{fig:b-T}(a) and \ref{fig:b-T}(b). 
In Fig. \ref{fig:b-T}(a), the synapses were not depressed, i.e, $\gamma=0.0$, while 
Fig. \ref{fig:b-T}(b) is the case with synaptic depression at $\gamma=0.5$.
The thick solid, thin solid, and dashed lines indicate the transition temperatures to the paramagnetic, 
memory state, and mixed state phase, respectively. 
First, we discuss the common feature of Figs. \ref{fig:b-T}(a) and \ref{fig:b-T}(b). 
When correlation $b$ was sufficiently large, 
only the mixed state was stable
because the distances between memory patterns was small.
On the other hand, with small $b$, 
both the memory and mixed states could be stable. 
In the high temperature region, the phase became paramagnetic. 
As shown in Fig. \ref{fig:b-T}(b),
regardless of the correlation coefficient, the unstable phase can exist at intermediate temperatures 
as a result of synaptic depression. \\
In Fig. \ref{fig:b-T}, $b_c$ means the maximum value of correlation coefficient $b$ 
for which the memory state can be stable.
We have analytically proved (see the appendix for details) that $b_c$ 
has the same value, $1/\sqrt{2}$, regardless of depression parameter $\gamma$. 
Therefore, synaptic depression does not influence the stability of the memory state 
in $p\sim O(1)$ and $\beta \rightarrow \infty$. 

\subsection{Macroscopic behavior in unstable phase}
In the previous section, we described our investigation of the macroscopic steady state and 
found the unstable phase for various parameter values. 
Previous studies have reported that the network can be unstable 
owing to synaptic depression\cite{Pantic}. 
However, the dynamics of the macroscopic state in that phase 
were not demonstrated by a dynamical equation obtained by mean-field analysis. 
In the present study, we investigated 
how the behavior of the macroscopic state in the unstable phase by theory and computer simulation.\\ 
First, we assumed that neurons within the same sublattice $\mathcal{I}_{\textrm{\boldmath $\eta$}}$ 
follow the same dynamics. Therefore, the firing rate and the dynamic amplitude factor can be described as
\begin{equation}
m_i(t)=m_{\textrm{\boldmath $\eta$}}(t),\: X_i(t)=X_{\textrm{\boldmath $\eta$}}(t), \: i\in\mathcal{I}_{\textrm{\boldmath $\eta$}}.
\label{eq:assu_osci}
\end{equation}
Under this assumption, from eqs. (\ref{eq:s_mean_dym}) and (\ref{eq:x_mean_dym}), 
we obtained macroscopic dynamical equations for the sublattice mode:
\begin{eqnarray}
\label{eq:m_sub_dym}
m_{\textrm{\boldmath $\eta$}}(t+1) 
&=&
g_{\beta}( \langle h_{\textrm{\boldmath $\eta$}}(t) \rangle ), \\
\langle h_{\textrm{\boldmath $\eta$}}(t)\rangle&=&\sum_{\textrm{\boldmath $\eta$}'}p_{\textrm{\boldmath $\eta$}'}\textrm{\boldmath $\eta$}
\cdot\textrm{\boldmath $\eta '$}(2m_{\textrm{\boldmath $\eta$}'}(t)X_{\textrm{\boldmath $\eta$}'}(t)-1),\\
\lefteqn{}X_{\textrm{\boldmath $\eta$}}(t+1)  &=&
X_{\textrm{\boldmath $\eta$}}(t)+\frac{1-X_{\textrm{\boldmath $\eta$}}(t)}{\tau}-
Um_{\textrm{\boldmath $\eta$}}(t)X_{\textrm{\boldmath $\eta$}}(t).
\label{eq:x_sub_dym}
\end{eqnarray}
For simplicity, the initial state was set to the first memory pattern $\textrm{\boldmath $\xi$}^1$ 
in the case without synaptic depression 
as described below. 
\begin{eqnarray}
&&M^1(0)=1.0,\:\:M^2(0)=M^3(0)=b^2,\\
&&s_{i}(0)=
\left\{
\begin{array}{ll}
 1,\:\:(\xi^{1}_i=1), &\quad  \\
 0,\:\:(\xi^{1}_i=-1),&\quad
\end{array}
\right.\\
&&x_i=1.0.
\end{eqnarray}
These equations correspond to the following condition in terms of sublattice mode. 
\begin{eqnarray}
&&m_{\textrm{\boldmath $\eta$}}(0)=
\left\{
\begin{array}{ll}
 1.0,\:\:(\eta^{1}=1), &\quad  \\
 0.0,\:\:(\eta^{1}=-1), &\quad
\end{array}
\right.\\
&&X_{\textrm{\boldmath $\eta$}}(0)=1.0.
\end{eqnarray}
Figures \ref{fig:osci}(a)-(c) show the dependence of the overlaps $M^{\mu}(t)\:(\mu=1,2,3)$ 
defined by eq. (\ref{eq:overlap}) on time $t$ in the unstable phase. 
They show that different oscillatory behavior occurred depending on parameters $b$ and $T$.
Here, time $t$ corresponds to the Monte Carlo step of the computer simulation. 
The solid lines represent the dependence of $M^1$ on time $t$ 
obtained by numerically solving eqs. (\ref{eq:m_sub_dym})-(\ref{eq:x_sub_dym}). 
The dotted and dashed lines similarly represent those of $M^2$ and $M^3$, respectively. 
The black dots in the top panels indicate $M^1(t)$ obtained by computer simulation with $N=9.6 \times 10^4$. 
The eight bottom panels in each column show the dependences of 
$m_{\textrm{\boldmath $\eta$}}$ and $X_{\textrm{\boldmath $\eta$}}$ on time $t$ obtained by a theoretical approach. 
The thick and thin lines in these figures correspond to $m_{\textrm{\boldmath $\eta$}}$ and 
$X_{\textrm{\boldmath $\eta$}}$, respectively. 
Figures \ref{fig:osci}(a) and \ref{fig:osci}(b) show good agreement between the theory represented by lines and computer simulation represented by black dots. 
However, we can see that the theoretical and simulation results are out of phase in Fig. \ref{fig:osci}(c). 
This is because the assumption that neurons within same sublattice $\mathcal{I}_{\textrm{\boldmath $\eta$}}$ 
follow the same dynamics is not strictly valid. 
However, the periods of the dynamics derived from theory and computer simulation do coincide.
\\ 
The overlap $M^1$ periodically oscillated as shown in Fig. \ref{fig:osci}(a) 
when correlation coefficient $b$ was very small, i.e., $b=0.05$ and $T=0.65$. 
In this case, $M^2$ and $M^3$ also slightly oscillated keeping the same values as each other.
That is to say, the network switched between the first memory pattern $\textrm{\boldmath $\xi$}^1$ 
and the anti memory pattern $-\textrm{\boldmath $\xi$}^1$ in cycles. 
We see in the bottom figures that $m_{\textrm{\boldmath $\eta$}}$ and $X_{\textrm{\boldmath $\eta$}}$ oscillated 
with the same phase when 
the first components of sublattice indices $\eta^1$ had the same value. \\
When the distances between stored patterns was sufficiently small, which corresponds to the case of large $b$, 
each overlap oscillated with the same value at $b=0.8$ and $T=1.4$, 
as shown in Fig. \ref{fig:osci}(b). 
In this case, the network periodically oscillated between the mixed and anti-mixed states. 
The bottom figures in Fig. \ref{fig:osci}(b), illustrate that 
$m_{\textrm{\boldmath $\eta$}}$ and $X_{\textrm{\boldmath $\eta$}}$ 
oscillated with the same phase when $\eta^1 +\eta^2 +\eta^3$ had the same value. 
This is attributed to the mixed state, $\textrm{sgn}(\eta^1 +\eta^2 +\eta^3)$. \\
At $b=0.35$ and $T=0.5$ (Fig. \ref{fig:osci}(c)), unlike both 
the macroscopic behavior shown in Figs. \ref{fig:osci}(a) and \ref{fig:osci}(b), 
the overlaps oscillated while keeping a positive value, i.e., 
$M^{\mu}>0$, and switched to each other. 
In other words, the network toured all the memory patterns in turn. 
We examined the oscillation at $b=0.35$ and $T=0.5$ in more detail. 
Figure \ref{fig:osci_2} illustrates the oscillation represented in Fig. \ref{fig:osci}(c) 
within the phase space composed of $(M^1,M^2,M^3)$ obtained from theory and simulation. 
The phase space is represented as a two-dimensional space whose plane surface contains all the attractors. 
The apexes of the triangles correspond to the coordinates of the memory patterns 
for eqs. (\ref{eq:overlap_pattern}) and (\ref{eq:overlap}), 
i.e., $(1,b^2,b^2)$, $(b^2,1,b^2)$ and $(b^2,b^2,1)$. 
The center dots in the triangles indicate the coordinates of 
the mixed state in phase space. 
The solid lines in the three triangles represent the locus of the state in a period, 
and the three figures at the top and bottom in Fig. \ref{fig:osci_2} are in chronological order from left to right. 
We can see that the macroscopic state starting 
from the first memory pattern $\textrm{\boldmath $\xi$}^1$ did not stay in the attracter. 
Consequently, in such a cyclic behavior, the state gravitated toward the mixed state once, but 
it could not remain there and immediately transited to another memory pattern. 

\section{Conclusion}
In this paper, 
we discussed the associative memory model with synaptic depression 
and applied the dynamical mean-field theory with the notion of a sublattice to the model 
by a statistical mechanical approach. 
We then considered a model
that stores three correlated memory patterns,
and examined how the stability of each steady state can change depending on the strength of synaptic depression 
and the correlation level among the memory patterns.
Our theory enables us to treat 
the stability of not only the memory state but also the mixed state.  
As a result, we found that there is an unstable phase in which the network 
could not remain in any attractors.  
Furthermore, we investigated the macroscopic dynamics in the unstable phase 
and showed that three different types of oscillation existed in that phase 
depending on certain parameters: the fraction of synaptic depression, recovery time constant, 
and correlation level among the memory patterns. 
The first one is the oscillation between the memory and anti-memory states: 
this oscillation has been reported in previous work\cite{Pantic}. 
The second is 
the oscillation between the mixed and anti-mixed states, which
occurred owing to the correlation among memory patterns. 
The third is the oscillation in which 
the network toured among the memory patterns periodically via the mixed state. 
This switching phenomenon may be connected with a search among similar memories and ``attractor ruins" in terms of chaos. \cite{Marro_1}\cite{Tsuda}\\
In this study, we focused on a small number of stored patterns. 
In future work, we compute the starage capacity of the associative memory model with synaptic depression
where we consider the case in which the number of stored patterns is on the order of $N$.
\appendix

\section{Stability of memory pattern at the low temperature limit}
At the low temperature limit ($\beta \rightarrow \infty$), the stability of the memory state has nothing to do with 
synaptic depression as shown by Fig.{\ref{fig:b-T}}.
We analytically prove this statement in this section.\\
We rewrite the overlap (\ref{eq:overlap}) in the steady state as
\begin{eqnarray}
\lefteqn{}M^{\mu} &=&\frac{1}{N}\sum_{i}^N \xi_i^{\mu} (2s_i-1), \\
 &\rightarrow&\frac{1}{N}\sum_{i}^N \xi_i^{\mu}\mathrm{sgn}(h_i)\:(\beta\rightarrow \infty),
 \label{eq:M_ape}
\end{eqnarray}
where we use eq. (\ref{eq:glauber_dym}). 
By considering the dynamic amplitude factor in the steady state $(x_j=1/1+\gamma s_j)$, 
we can denote the internal potential $h_i$ of the $i$-th neuron as 
\begin{eqnarray}
\lefteqn{}h_i &=&\frac{1}{N}\sum_{j\neq i}\sum_{\mu=1}^{p}\xi_{i}^{\mu}\xi_{j}^{\mu}(2s_jx_j-1), \\
&\sim&\frac{1}{N}\sum_{j=1}^N\sum_{\mu=1}^{p}\xi_{i}^{\mu}\xi_{j}^{\mu}(2s_jx_j-1),\\
 & =&\frac{1}{N}\sum_{j=1}^N\sum_{\mu=1}^{p}\xi_{i}^{\mu}\xi_{j}^{\mu}\left(2\frac{s_j}{1+\gamma s_j}-1 \right).
\label{eq:h_ape}
\end{eqnarray}
by using eqs. (\ref{eq:h_original}) and (\ref{eq:interaction}) for $N\rightarrow \infty$ and $p\sim(1)$.
Here, we use the following reasonable approximation at $\beta\rightarrow \infty$ \cite{Matsumoto}.
\begin{equation}
\frac{s_j}{1+\gamma s_j}=\frac{s_j}{1+\gamma }.
\end{equation}
Then, eq. (\ref{eq:h_ape}) is rewritten as
\begin{eqnarray}
\lefteqn{} h_i&=& \frac{1}{N}\sum_{\mu=1}^{p}\xi_{i}^{\mu}\sum_{j=1}^{N}\xi_j^{\mu} \left(\frac{2s_j}{1+\gamma}-1 \right),\\ \nonumber
 &= &\frac{1}{N(1+\gamma)}\sum^{p}_{\mu=1}\xi_{i}^{\mu}\sum_{j=1}^{N}\xi_j^{\mu} \left(2s_j-1 \right)\\ 
&&- \frac{\gamma}{N(1+\gamma)}\sum^{p}_{\mu=1}\xi_{i}^{\mu}\sum_{j=1}^{N}\xi_j^{\mu}, \\ 
 &= & \frac{1}{1+\gamma}\sum_{\mu=1}^{p}\xi_i^{\mu}M^{\mu},
 \label{eq:h_limit}
\end{eqnarray}
where we use $\frac{1}{N}\sum_{j=1}^{N} \xi_j^{\mu}=0\:(N\rightarrow \infty)$ for eq. (\ref{eq:xi_prob}) in the last step. 
Therefore, for $0<\gamma$ and eq. (\ref{eq:M_ape}), the overlap $M^{\mu}$ is written as
\begin{equation}
M^{\mu}=\langle  \langle \mathrm{sgn}(\sum_{\mu=1}^{p}\xi_i^{\mu}M^{\mu}) \rangle \rangle,
\label{eq:M_low}
\end{equation}
where $\langle \langle \cdot \rangle \rangle$ represents the average 
with respect to stochastic variable $\textrm{\boldmath $\xi$}^{\mu}$.
This is equal to the overlap in the case without synaptic depression at the low temperature limit ($\beta \rightarrow \infty$).
When the state is the first memory pattern $\textrm{\boldmath $\xi$}^1$, i.e., $(M^1,M^2,M^3)=(1,b^2,b^2)$, for example, eq. (\ref{eq:M_low}) becomes
\begin{equation}
1=\langle  \langle \mathrm{sgn}(1+b^2\sum^{3}_{\nu=2}\xi_i^{\nu}) \rangle \rangle.
\end{equation}
Consequently, in the condition $b<1/\sqrt{2}=b_c$, the memory 
pattern phase ``ME" is stable regardless of synaptic depression.


\begin{thebibliography}{99}

\bibitem{Nakano}
K. Nakano: IEEE Transactions on Systems, Man and Cybernetics \textbf{2} (1972) 381.

\bibitem{Anderson}
J. A. Anderson: Mathmatical Biosciences \textbf{14} (1972) 197.

\bibitem{Kohonen}
T. Kohonen: IEEE Transactions on Computers \textbf{21} (1972) 353.

\bibitem{Hopfield}
J. J. Hopfield: Proc. Natl. Acad. Sci. U.S.A.  \textbf{79} (1982) 2554.

\bibitem{Mimura}
K. Mimura: J. Phys. Soc. Jpn \textbf{78} (2009) 033001.

\bibitem{Pantic}
L. Pantic, J. J. Torres, H. Kappen, and S. Gielen: Neural Comput. \textbf{14} (2002) 2903. 

\bibitem{Marro_1}
J. Marro, J. J. Torres and J. M. Cortes: Neural Networks \textbf{20} (2007) 230. 

\bibitem{Torres_2}
J. J. Torres, J. Marro J. M. Cortes and B. Wemmenhove: Neural Networks \textbf{21} (2008) 1272. 

\bibitem{Melamed}
O. Melamed, O. Barak, G. Silberberg, H. Markram and M. Tsodyks: J.Compt Neurosci \textbf{25} (2008) 308. 

\bibitem{Sompolinsky}
H. Sompolimsky and I. Kanter: Phys. Rev. Lett. \textbf{57} (1986) 2861 

\bibitem{Tsuda}
I.Tsuda, E. Koerner and H. Shimizu: Prog. Theor. Phys. \textbf{78} (1987) 51.

\bibitem{Amit}
D. J. Amit: Proc. Nati. Acad. Sci. USA \textbf{85} (1988) 2141. 

\bibitem{Fukai}
T. Fukai and M. Shiino: Phys. Rev. Lett. \textbf{64} (1990) 1465

\bibitem{Thomson}
A. Thomson and J. Deuchars: Trends Neurosci. \textbf{17} (1994) 119.

\bibitem{Abbott}
L. Abbott, J. Varela, K. Sen, and S. Nelson: Science \textbf{275} (1997) 220.

\bibitem{Tsodyks}
M. Tsodyks and H. Markram: Proc. Natl. Acad. Sci. U.S.A. \textbf{94} (1997) 719.

\bibitem{Amari}
S. Amari: Biol. Cybern. \textbf{26} (1977) 175.

\bibitem{Toya}
K. Toya, K. Fukushima, Y. Kabashima, and M. Okada: J. Phys. A \textbf{33} (2000) 2725.

\bibitem{Igarashi_0}
Y. Igarashi, M. Oizumi, Y. Otsubo, K. Nagata and M. Okada: J. Phys. Conf. Ser. \textbf{197} (2009) 012018.

\bibitem{Igarashi}
Y. Igarashi, M. Oizumi and M. Okada: arXiv:1003.1196.

\bibitem{Matsumoto}
N. Matsumoto, D. Ide, M. Watanebe, and M. Okada: J. Phys. Soc. Jpn \textbf{76} (2007) 084005.

\bibitem{Mejias}
J. F. Mejias and J. Torres: Neural Comput \textbf{21} (2009) 851.

\bibitem{Torres}
J. Torres, L. Pantic, and H. Kappen: Phys. Rev. E \textbf{66} (2002) 061910.

\bibitem{Coolen}
A. C. C. Coolen: cond-mat/0006011.


\end{thebibliography}
\end{document}